\documentstyle[aps,psfig]{revtex} 
\newcommand{\be}{\begin{equation}} 
\newcommand{\ee}{\end{equation}} 
\newcommand{\ben}{\begin{eqnarray}} 
\newcommand{\een}{\end{eqnarray}} 
\begin{document} 
 
\twocolumn[\hsize\textwidth\columnwidth\hsize\csname 
@twocolumnfalse\endcsname 
 
\title{Tiling the plane without supersymmetry} 
\author{D. Bazeia and F.A. Brito} 
\address{Departamento de F\'\i sica, Universidade Federal da Para\'\i ba,\\ 
Caixa Postal 5008, 58051-970 Jo\~ao Pessoa, Para\'\i ba, Brazil} 
\date{\today} 
 
\maketitle 
 
\begin{abstract} 
We present a way of tiling the plane with a regular hexagonal network 
of defects. The network is stable and follows in consequence of the 
three-junctions that appear in a model of two real scalar fields 
that presents $Z_3$ symmetry. The $Z_3$ symmetry is effective 
in both the vacuum and defect sectors, and no supersymmetry 
is required to build the network.  
\end{abstract}  
\vskip2pc] 
  
\newpage 
  
Domain walls appear in diverse branches of physics, envolving energy scales 
as different as the ones for instance in magnetic materials \cite{esc81} 
and in cosmology \cite{vsh96}. They live in three 
spatial dimensions as bidimensional objects that arise in systems 
with at least two isolated degenerate minima. In field theory they 
appear in the $(3,1)$ dimensional space-time, and this may happen in
supersymmetric theories, although supersymmetry plays no fundamental 
role for the presence of domain walls. 
 
Very recently, in a paper by Gibbons and Townsend \cite{gto99}, 
and also in Refs.~{\cite{cht99,saf99}}, one investigates the presence
of domain walls and their possible intersections in a Wess-Zumino model, 
with a polynomial superpotential. In the supersymmetric theory, one can
classify the classical solutions as BPS and non-BPS states, according 
to the work of Bogomol'nyi, and of Prasad and Sommerfield \cite{bps75}. 
The BPS states are stable, and are expected to play some role
in investigating duality in supersymmetric models. We recall
that no BPS state can be annihilated under continuum variation of
the parameters that define the supersymmetric theory. 

In Ref.~{\cite{b95}} one investigates models
of coupled real scalar fields in bidimensional space-time. These  
investigations provide a concrete way of finding BPS states and suggest  
other studies, in particular on the subject of defects inside  
defects -- see Ref.~{\cite{b96}}. Most of the  
models investigated in \cite{b95,b96} can be seen as real bosonic portions  
of supersymmetric theories. In supersymmetric models the presence of discrete
symmetry may produce BPS and non-BPS defects. The BPS states lie in shorter
multiplets, and preserve the supersymmetry only partially \cite{wol78,dsh97}.  
There are BPS states that preserve $1/2$ of the supersymmetry, but the 
possibility of BPS states preserving $1/4$ supersymmetry is subtler,  
and is shown to appear as junctions \cite{cqr91,ato91} of domain walls  
in the recent papers \cite{gto99,cht99,saf99}. 
 
In the present work we start dealing with the bosonic portions  
of supersymmetric theories. We do this guided by the discrete $Z_3$  
symmetry, with the aim of describing the presence of  
three-junctions and the network of defects that it can generate.  
We first point out that supersymmetry introduces restrictions
that may lead to instability of the junction, or at least of the
network that it could generate. We then examine another model,
and show that all the difficulties found in the supersymmetric
context are circumvented by just giving up supersymmetry.

The subject of this work may be of interest to several different branches
of physics, in particular in applications concerning the entrapment of
networks of defects inside domain walls. This possibility can be implemented
with three scalar fields, in models engendering the $Z_2\times Z_3$ symmetry,
following the lines of Refs.~{\cite{b96,bbr99b}}. Other applications may
include strong interactions, if we recall that the $Z_3$ group is the center
of the $SU(3)$ group -- see for instance Refs.~{\cite{ctr98,chw98}}. Also,
there are applications to systems of condensed matter, in particular on
issues concerning pattern formation \cite{wal97}, as for instance in the
case of the thermal convection studied in Ref.~{\cite{ccl90}}.

We start describing the two real scalar fields $\phi$ 
and $\chi$ in bidimensional space-time with the Lagrangian density  
\be  
\label{L}  
{\cal L}=\frac{1}{2}\partial_{\alpha}\phi\partial^{\alpha}\phi+  
\frac{1}{2}\partial_{\alpha}\chi\partial^{\alpha}\chi-V  
\ee  
Here $V=V(\phi,\chi)$ is the potential. In the supersymmetric case 
it has the general form  
\be  
\label{ssys}  
V(\phi,\chi)=\frac{1}{2}\,W^2_{\phi}+\frac{1}{2}\,W^2_{\chi}  
\ee 
where $W=W(\phi,\chi)$ is the superpotential.  
 
The superpotential allows introducing several properties,  
as shown in Refs.~{\cite{b95,b96}}. For instance, for static  
fields the equations of motion are solved by first-order 
differential equations $d\phi/dx=W_{\phi}$ and
$d\chi/dx=W_{\chi}$. The energy of solutions of the
first-order equations are given by $E^{ij}_B=|\Delta W_{ij}|$,
with $\Delta W_{ij}=W_i-W_j$ and $W_i=W(\phi_i,\chi_i)$, where $(\phi_i,\chi_i)$ represents the i-th vacuum state of the model.
This is the Bogomol'nyi bound, and the corresponding solutions are
BPS solutions. The BPS solutions are linearly or classically stable.

We guide ourselves toward the topological solutions by introducing the 
topological current
\be
\label{cur}
J^{\alpha}=\varepsilon^{\alpha\beta}\partial_{\beta}{\phi\choose\chi}
\ee
It obeys $\partial_{\alpha}J^{\alpha}=0$, and it is also a vector 
in the $(\phi,\chi)$ plane. For static configurations we have 
$J^{t}_{\alpha}\,J^{\alpha}=\rho^t\,\rho$, where $\rho=\rho(\phi,\chi)$ 
is the charge density. This charge density allows writing $\rho^t\rho$ 
as twice the kinetic energy density of the topological solution,
and this can be used to infer stability of junctions -- see below. 

Let us now consider a specific model, defined by the 
superpotential \cite{bbr99b} 
\ben  
W(\phi,\chi)&=&\lambda\phi^3-3\lambda\phi+\lambda\phi\left(\chi+ 
\sqrt{\frac{1}{3}}\right)^2\nonumber  
\\  
& &+\lambda\sqrt{\frac{4}{27}}\left(\chi+\sqrt{\frac{1}{3}}\right)^3-  
\lambda\left(\chi+\sqrt{\frac{1}{3}}\right)^2  
\een 
The first-order differential equations are given by  
\ben  
\frac{d\phi}{dx}&=&3\lambda (\phi^2-1)+\lambda\left(\chi+ 
\sqrt{\frac{1}{3}}\right)^2  
\\  
\frac{d\chi}{dx}&=&2\lambda\left(\chi+\sqrt{\frac{1}{3}}\right)  
\Biggl[\phi+\sqrt{\frac{1}{3}}\left(\chi+\sqrt{\frac{1}{3}}\right)  
-1\Biggr] 
\een 
The model is described by the fourth-order polynomial potential 
\ben 
V_4(\phi,\chi)&=&\frac{9}{2}\lambda^2(\phi^2-1)^2+ 
\frac{1}{2}\lambda^2\left(\chi+\sqrt{\frac{1}{3}}\right)^4\nonumber 
\\ 
& &+3\lambda^2(\phi^2-1)\left(\chi+\sqrt{\frac{1}{3}}\right)^2\nonumber 
\\  
& &+2\lambda^2\left(\chi+\sqrt{\frac{1}{3}}\right)^2\Biggl[\phi+  
\sqrt{\frac{1}{3}}\chi-\frac{2}{3}\Biggr]^2  
\een  
Thus, it behaves standardly in one, two, and three spatial dimensions. 
 
The above potential presents the three vacuum states $(0,\sqrt{4/3}),\, 
(-1,-\,\sqrt{1/3})$, and $(1,-\,\sqrt{1/3})$. These minima form an 
equilateral triangle invariant under the $Z_3$ rotations, with side 
$l=2$. The values of the superpotential at the minima are 
$W_1=-\lambda,\, W_2=2\,\lambda$ and $W_3=-2\lambda$. The energies 
of the BPS states are $|\lambda|,\,3\,|\lambda|,$ and $4\,|\lambda|$. 
All the three sectors are BPS sectors, but they do not present the $Z_3$ 
symmetry that connects the vacuum states. The highest energy is 
associated to the sector connecting the second and third vacuum states. 
This is the only sector where we can find explicit solutions. 
They are given by $\phi(x)=-\tanh(3\lambda\,x)$ and $\chi(x)=-\sqrt{1/3\,}$.  
This is a BPS state, representing an orbit in the  
$(\phi,\chi)$ plane. The orbit is a straight-line segment that 
connects the corresponding vacuum states. The orbits connecting 
the other vacua cannot be straight-line segments. They cannot be  
obtained by rotating the $(\phi,\chi)$ plane according to the  
$Z_3$ symmetry, and so the defect sectors do not present the  
$Z_3$ symmetry that connects the vacuum states. This fact also  
appears when one identifies the tensions of the BPS defects.  
They are given by $t_1=|\lambda|$, $t_2=3\,|\lambda|$, and  
$t_3=4\,|\lambda|$. They are different and do not obey the $Z_3$ 
symmetry. They are such that $t_3=t_1+t_2$, and do not strictly obey  
the triangle inequality one needs to ensure stability \cite{ato91} 
of the three-junction that appears in this model. 

To circumvent instability of the three-junction we now follow 
Refs.~{\cite{gto99,cht99}}.  We make contact with these works 
after considering superpotentials that satisfy $W_{\phi\phi}+W_{\chi\chi}=0$.  
In this case, for harmonic superpotentials one adds to the two  
first-order equations $d\phi/dx=W_{\phi}$ and $d\chi/dx=W_{\chi}$ the
two new \cite{bbr99b} first-order equations: $d\phi/dx=-W_{\chi}$ 
and $d\chi/dx=W_{\phi}$. Solutions to these equations also 
minimize the energy and solve the equations of motion. This allows 
introducing ${\widetilde W}(\phi,\chi)$ such that 
${\widetilde W}_{\phi}=-W_{\chi}$ and ${\widetilde W}_{\chi}=W_{\phi}$. 
We use $W$ and ${\widetilde W}$ to introduce the complex superpotential, 
${\cal W}=W+i\,{\widetilde W}$. We write the complex superpotential 
in terms of the complex field $\phi+i\chi$, and this is the way one 
gets from the investigations of Refs.~{\cite{b95,b96}} to the recent 
possibility \cite{gto99,cht99,saf99} of describing three-junctions 
preserving $1/4$ supersymmetry. However, junctions require 
the presence of at least three minima, and this is only achieved when 
the superpotential is of at least the fourth-order power in the complex 
field. This means that the model behaves standardly only in one and two 
spatial dimensions. In this case one can show explicitly \cite{gto99,cht99} 
that the three-junction is stable and breaks $1/4$ supersymmetry, although 
supersymmetry itself does not allow the presence of a stable network of 
defects \cite{gto99,cht99,saf99}. Owing to the fact that each adjacent 
junction in the network has opposite winding number, any adjacent vacua 
should be connected with defect solutions also having opposite winding 
numbers along the same orbit \cite{saf99}. Since we have to use different 
conjugate Bogomol'nyi equations to take into account these winding 
numbers, the network cleary cannot be BPS and then can decay. 
 
We then give up supersymmetry, turning our attention to polynomial 
potentials that engenders the $Z_3$ symmetry, and that supports 
stable three-junctions that generate a regular hexagonal network of defects. 
Interestingly, we have found a fourth-order polynomial potential that do the 
job. It is given by 
\ben 
\label{p}  
V(\phi,\chi)&=&\lambda^2\phi^2\left(\phi^2-\frac{9}{4}\right)+ 
\lambda^2\chi^2\left(\chi^2-\frac{9}{4}\right)\nonumber  
\\ 
& &+\,2\,\lambda^2\,\phi^2\,\chi^2-\lambda^2\phi\,(\phi^2-3\,\chi^2)+ 
\frac{27}{8}\lambda^2  
\een  
\label{sm1}  
This potential was introduced in Ref.~{\cite{ruc80}}. The equations 
of motion for static configurations are  
\ben  
\label{sm21}  
\frac{d^2\phi}{dx^2}&=&\lambda^2\phi\left(4\phi^2+ 
4\chi^2-3\,\phi-\frac{9}{2}\right)+3\lambda^2\chi^2  
\\  
\label{sm22}  
\frac{d^2\chi}{dx^2}&=&\lambda^2\chi\left(4\phi^2+4\chi^2+ 
6\,\phi-\frac{9}{2}\right)  
\een 
The potential has three degenerate minima, at the points $v_1=(3/2)\,(1,0)$ 
and $v_{2,3}=(3/4)\,(-1,\pm\sqrt{3})$. These minima form 
an equilateral triangle, invariant under the $Z_3$ symmetry. 
The distance between the minima is $(3/2)\sqrt{3}$. 
 
We can obtain the topological solutions explicitly. The easiest way 
to do this follows by first examining the sector that connects the vacua 
$v_2$ and $v_3$. This is so because 
in this case we set $\phi=-3/4$, searching for a strainght-line 
segment in the $(\phi,\chi)$ plane. This is compatible with the 
Eq.~(\ref{sm21}), and reduces the other Eq.~(\ref{sm22}) to the form 
\be 
\frac{d^2\chi}{dx^2}=\lambda^2\left(4\chi^3-\frac{27}{4}\chi\right) 
\ee 
This implies that the orbit connecting the vacua $v_2$ and $v_3$ 
is a straight line. It is such that, along the orbit the 
$\chi$ field feels the potential $\lambda^2\,[\chi^2-(27/16)]^2$. 
This shows that the model reduces to a model of a single field, 
and the solution satisfies the first-order equation 
\be 
\frac{d\chi}{dx}=\sqrt{2}\lambda\left(\chi^2-\frac{27}{16}\right) 
\ee 
The solution is 
\be 
\chi(x)=-\frac{3}{4}\sqrt{3}\tanh\left(\sqrt{\frac{27}{8}}\lambda\,x\right) 
\ee 
The other solutions can be obtained by rotations obeying the $Z_3$ symmetry 
of the model. 
 
The full set of solutions of the equations of motion are collected below.  
In the sector connecting the minima $v_2$ and $v_3$ they are  
\ben  
\label{sol11}  
\phi^{(\pm)}_{(2,3)}&=&-\frac{3}{4}  
\\  
\label{sol12}  
\chi^{(\pm)}_{(2,3)}&=&\pm\frac{3}{4}\sqrt{3}\, 
\tanh\left(\sqrt{\frac{27}{8}}\, \lambda\,x\right)  
\een  
In the sector connecting the minima $v_1$ and $v_2$ they are  
\ben  
\label{sol21}  
\phi^{(\pm)}_{(1,2)}&=&\frac{3}{8}  
\pm\frac{9}{8}\,\tanh\left(\sqrt{\frac{27}{8}}\,\lambda\,x\right)  
\\  
\label{sol22}  
\chi^{(\pm)}_{(1,2)}&=&\frac{3}{8}\sqrt{3}  
\mp\frac{3}{8}\sqrt{3}\,\tanh\left(\sqrt{\frac{27}{8}}\,\lambda\,x\right)  
\een  
In the sector connecting the minima $v_1$ and $v_3$ they are  
\ben  
\label{sol31}  
\phi^{(\pm)}_{(1,3)}&=&\frac{3}{8}  
\mp\frac{9}{8}\,\tanh\left(\sqrt{\frac{27}{8}}\,\lambda\,x\right)  
\\  
\label{sol32}  
\chi^{(\pm)}_{(1,3)}&=&-\frac{3}{8}\sqrt{3}  
\mp\frac{3}{8}\sqrt{3}\,\tanh\left(\sqrt{\frac{27}{8}}\,\lambda\,x\right)  
\een  
The label $(\pm)$ is used to identify kink and antikink. All the solutions  
have the same energy, $(9/4)\,\sqrt{27/8}\,|\lambda|$. 
 
We examine how the bosonic fields behave in the background 
of the classical solutions. We do this by considering fluctuations 
around the static solutions $\phi(x)$ and $\chi(x)$. 
We use the equations of motion to see that the fluctuations 
depend on the potential 
\be 
{\bf U}(x)={ {V_{\phi\phi}\,\,\, V_{\phi\chi} } 
\choose{ V_{\chi\phi}\,\,\, V_{\chi\chi} } } 
\ee 
Evidently, after obtaining the derivatives we substitute 
the fields by their classical static values $\phi(x)$ 
and $\chi(x)$. The model under consideration is defined 
by the potential (\ref{p}). In this case we use (\ref{sol11}) 
and (\ref{sol12}) to obtain two decoupled equations for the 
fluctuations. The potentials of the corresponding Schr\"odinger-like 
equations are 
\ben 
U_{11}(x)&=&\frac{27}{8}\lambda^2\Biggl[4-2\, 
{\rm sech}^2\left(\sqrt{\frac{27}{8}}\lambda\,x\right)\Biggr] 
\\ 
U_{22}(x)&=&\frac{27}{8}\lambda^2\Biggl[4-6\, 
{\rm sech}^2\left(\sqrt{\frac{27}{8}}\lambda\,x\right)\Biggr] 
\een 
The eigenvalues can be obtained explicitly: in the $\chi$ direction 
we get $w^{\chi}_0=0$ and $w^{\chi}_1=(9/2)\sqrt{\lambda^2/2}$, 
and in the $\phi$ direction we have $w^{\phi}_0=(9/2)\sqrt{\lambda^2/2}$. 
This shows that the pair $(\ref{sol11})$ and $(\ref{sol12})$ 
is stable, and by symmetry we get that all the three topological 
solutions are stable solutions. 
 
The classical solutions present the nice property of having energy 
evenly distributed in their kinetic (k) and potential (p) 
portions. In terms of energy density they are 
\be 
{\rm k}(x)={\rm p}(x)=\frac{1}{4}\left(\frac{27}{8}\right)^2\,\lambda^2\,
{\rm sech}^{4}\left(\sqrt{\frac{27}{8}}\lambda\,x\right) 
\ee 
To understand this feature we recall the calculation done 
explicitly in the sector with $\phi=-3/4$, constant. There the model 
is shown to reduce to a model of a single field, a model that 
supports BPS solutions. Within this context, the above 
solutions are very much like the non-BPS solutions that appear in 
supersymmetric systems \cite{bbr99b}. We use this property and the 
topological current (\ref{cur}) to obtain $\rho^t\rho=\varepsilon$, 
where $\varepsilon(x)={\rm k}(x)+{\rm p}(x)$ is the (total) energy 
density of the solution. We use this result and the notation $ij$, 
to identify the sector connecting the vacua $(\phi_i,\chi_i)$ and
$(\phi_j,\chi_j)$, to show that for any two different sectors 
$ij$ and $jk$, $i,j,k=1,2,3$ we get that
\be
(\rho_{ij}+\rho_{jk})^t\,(\rho_{ij}+\rho_{jk})< 
\rho_{ij}^t\rho_{ij}+\rho_{jk}^t\rho_{jk}
\ee
This condition shows that the three-junction is a process
of fusion of defects that occurs exothermically, providing
stability of junctions in the present model. This result
is more general than the one in Ref.~{\cite{ato91}}, which
appears within the context of supersymmetry. Evidently, our
result also works for BPS and non-BPS solutions that appears in
supersymmetric models, with the property of having energy evenly
distributed in their kinetic and potential portions \cite{bbr99b}. 
 
We notice that the orbits corresponding to the stable defect solutions  
form an equilateral triangle in the $(\phi,\chi)$ plane. This is so because 
the solutions are straight-line segments joining the three vacuum states  
in configuration space. They are degenerate in energy, and this allows
associating to each defect the same tension
\be
\label{ten}
t=\frac{9}{4}\,\sqrt{\frac{27}{8}}\,\,|\lambda|
\ee
This makes $t_{ij} < t_{jk}+t_{ki}, i,j,k=1,2,3$, and now the
inequality is strictly valid in this case, stabilizing the three-junction
that appears in this model when one enlarges the space-time to three
spatial dimensions.  
  
We consider the possibility of junctions in the plane, which may give  
rise to a planar network of defects. We work in the $(2,1)$ space-time,  
in the plane $(x,y)$. We identify the plane $(x,y)$ with the space of 
configurations, the plane $(\phi,\chi)$. We illustrate this situation by 
considering, for instance, the solutions  
we have already obtained. They are collected  
in Eqs.~(\ref{sol11})-(\ref{sol32}) in (1,1) dimensions.  
In the planar case they change to  
\ben  
\phi^{(\pm)}_{(2,3)}&=&-\frac{3}{4}  
\\  
\chi^{(\pm)}_{(2,3)}&=&\pm\frac{3}{4}\sqrt{3}  
\tanh\left(\sqrt{\frac{27}{8}}\,\lambda\,y\right)  
\een  
and  
\ben  
\phi^{(\pm)}_{(1,2)}&=&\frac{3}{8}\mp\frac{9}{8}  
\tanh\left(\frac{1}{2}\sqrt{\frac{27}{8}}\,\lambda\,(y+\sqrt{3}x)\right)  
\\  
\chi^{(\pm)}_{(1,2)}&=&\frac{3}{8}\sqrt{3}\pm\frac{3}{8}\sqrt{3}  
\tanh\left(\frac{1}{2}\sqrt{\frac{27}{8}}\,\lambda\,(y+\sqrt{3}x)\right)  
\een  
and  
\ben  
\phi^{(\pm)}_{(1,3)}&=&\frac{3}{8}\pm\frac{9}{8}  
\tanh\left(\frac{1}{2}\sqrt{\frac{27}{8}}\,\lambda\,(y-\sqrt{3}x)\right)  
\\  
\chi^{(\pm)}_{(1,3)}&=&-\frac{3}{8}\sqrt{3}\pm\frac{3}{8}\sqrt{3}  
\tanh\left(\frac{1}{2}\sqrt{\frac{27}{8}}\,\lambda\,(y-\sqrt{3}x)\right)
\een  
These planar defects are domain walls, and can be used to represent
the three-junction in the limit of thin walls.
 
\vspace{1.0cm} 
\begin{figure} 
\centerline{\psfig{figure=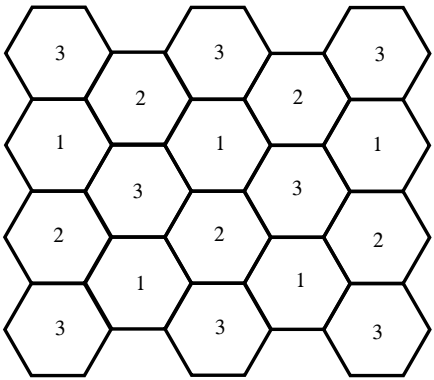,height=5.0cm}} 
\vspace{1.0cm} 
\caption{A regular hexagonal network of defects, formed by
three-junctions surrounded by domains representing the
vacua $v_1=1, v_2=2$, and $v_3=3$.} 
\end{figure}

The three-junction that appears in this $Z_3$-symmetric model allows  
building a network of defects, precisely in the form of a regular hexagonal  
network, as depicted in FIG.~1 in the thin wall approximation. In this
network the tension associated to the defect is the tipical value of the
energy in this tiling of the plane with a regular hexagonal network,
which seems to be the most efficient way of tiling the plane. As we
have shown, our model behaves standardly in $(3,1)$ dimensions.
It supports stable three-junctions that generate a stable regular
hexagonal network of defects.
 
We thank R.F. Ribeiro for discussions, and CAPES, CNPq, and PRONEX 
for partial support.

 
\end{document}